\documentclass[12pt]{article}
\usepackage{graphicx,amssymb,bbold}

\newcommand{\be}{\begin{equation}}
\newcommand{\ee}{\end{equation}}
\newcommand{\bea}{\begin{eqnarray}}
\newcommand{\eea}{\end{eqnarray}}
\newcommand{\beas}{\begin{eqnarray*}}
\newcommand{\eeas}{\end{eqnarray*}}

\begin{document}
\begin{titlepage}

\vspace*{-24mm}
\rightline{YITP-13-21}
\vspace{4mm}

\begin{center}

{\bf {\Large Black Hole Formation at the}} \\[3pt]
\vspace{1mm}
{\bf {\Large Correspondence Point}}

\vspace{8mm}

\renewcommand\thefootnote{\mbox{$\fnsymbol{footnote}$}}
Norihiro Iizuka${}^{1}$\footnote{iizuka@yukawa.kyoto-u.ac.jp},
Daniel Kabat${}^{2}$\footnote{daniel.kabat@lehman.cuny.edu},
Shubho Roy${}^{3,4}$\footnote{sroy@cts.iisc.ernet.in} and
Debajyoti Sarkar${}^{2,5}$\footnote{dsarkar@gc.cuny.edu}

\vspace{4mm}

${}^1${\small \sl Yukawa Institute for Theoretical Physics} \\
{\small \sl Kyoto University, Kyoto 606-8502, JAPAN}

${}^2${\small \sl Department of Physics and Astronomy} \\
{\small \sl Lehman College, City University of New York, Bronx NY 10468, USA}

${}^3${\small \sl Physics Department} \\
{\small \sl City College, City University of New York, New York NY 10031, USA}

${}^4${\small \sl Center for High Energy Physics} \\
{\small \sl Indian Institute of Science, Bangalore 560012, INDIA}

${}^5${\small \sl Graduate School and University Center} \\
{\small \sl City University of New York, New York NY 10036, USA}

\end{center}

\vspace{4mm}

\noindent
We study the process of bound state formation in a D-brane collision.
We consider two mechanisms for bound state formation.  The first,
operative at weak coupling in the worldvolume gauge theory, is pair
creation of W-bosons.  The second, operative at strong coupling,
corresponds to formation of a large black hole in the dual
supergravity.  These two processes agree qualitatively at intermediate
coupling, in accord with the correspondence principle of Horowitz and
Polchinski.  We show that the size of the bound state and timescale
for formation of a bound state agree at the correspondence point.  The
timescale involves matching a parametric resonance in the gauge theory
to a quasinormal mode in supergravity.

\end{titlepage}
\setcounter{footnote}{0}
\renewcommand\thefootnote{\mbox{\arabic{footnote}}}

\section{Introduction and summary}

Understanding black hole microstates from a D-brane or fundamental
string perspective is a long-standing theme in string theory.  The
original observation that vibrating strings qualitatively resemble a
black hole \cite{Susskind:1993ws,Sen:1995in} was followed by a
quantitative worldvolume derivation of black hole entropy for certain BPS states \cite{Strominger:1996sh}.  This
relationship eventually became a fundamental aspect of the holographic
duality between gauge and gravity degrees of freedom
\cite{Maldacena:1997re}.  According to this duality, microstates of a
black hole are in one-to-one correspondence with microstates of a
strongly-coupled gauge theory.  This duality also applies to
time-dependent processes such as black hole formation and evaporation,
leading to the viewpoint that these processes should be unitary,
contrary to \cite{Hawking:1976ra}.

To gain insight into black hole formation, and a better understanding
of the microstructure of the resulting black hole, in this paper we
study the process of bound state formation from two perspectives:
perturbative gauge theory and supergravity.  In perturbative gauge
theory a D-brane bound state can be formed through a process of open
string creation.  In supergravity we will see that open string
creation is not possible, and one instead forms a bound state through
the gravitational or closed-string process of black hole formation.

The perturbative gauge theory and supergravity calculations of bound
state formation do not have an overlapping range of validity.  But we
will show that they agree qualitatively at an intermediate value of
the coupling, in accord with the correspondence principle introduced
by Horowitz and Polchinski \cite{Horowitz:1996nw}.  This suggests that
there is a smooth transition between the process of open string
creation at weak coupling and black hole formation at strong coupling.

As a first test of these ideas, in \S \ref{D0collisions} we study
bound state formation in D0-brane collisions and show that the sizes
of the bound states match at the correspondence point.  In \S
\ref{Dpcollisions} we extend this analysis to general D$p$-branes.

Next we consider the time development of the bound states after they
have formed.  In \S \ref{sect:parametric} we show that the
weakly-coupled gauge theory has a parametric resonance which
exponentially amplifies the number of open strings present, and we
identify the timescale for the production of additional open strings
at weak coupling.  In the gravitational description, a perturbed black
hole approaches equilibrium on a timescale determined by the
quasinormal frequencies.  In \S \ref{comparison} we compare these two
timescales and show that they agree at the correspondence point.

In \S \ref{sect:equilibrium} we compare properties of the bound state
as initially formed to equilibrium properties of the black hole, and
show that at the correspondence point the bound state is created in a
state of near-equilibrium.  In \S \ref{gaugeshellcollapse} we study a
different initial configuration, in which a bound state is formed by
collapse of a spherical shell of D0-branes, and show that the picture
of a smooth transition between open string production and black hole
formation continues to hold.  We conclude in \S
\ref{sect:conclusions}.

The present work is related to several studies in the literature.  In
gauge -- gravity duality, a black hole on the gravity side is dual to
a thermal state of the gauge theory, where all ${\cal O}(N^2)$ degrees
of freedom are excited \cite{Witten:1998qj,Witten:1998zw}.  There have
been many studies of 0-brane black hole microstates from matrix
quantum mechanics, along with their associated thermalization process.
Some previous studies of 0-brane black holes from matrix quantum
mechanics include
\cite{Iizuka:2001cw,Kawahara:2007fn,Catterall:2007fp,Anagnostopoulos:2007fw,Hanada:2008gy,Hanada:2008ez,Catterall:2009xn}.
Also see \cite{Bhattacharyya:2009uu,Garfinkle:2011hm} for studies of
black hole formation from the gravity perspective, and
\cite{Festuccia:2006sa,Iizuka:2008hg,Iizuka:2008eb,Berenstein:2010bi,Asplund:2011qj,Asplund:2012tg}
for studies from the gauge theory perspective.  In particular
parametric resonance has been discussed in relation to thermalization
in the closely related work \cite{Berenstein:2010bi}.  Open string
production has been studied as a mechanism for trapping moduli at
enhanced symmetry points in \cite{Kofman:2004yc}, while open string
production in relativistic D-brane collisions has been studied in
\cite{McAllister:2004gd}.

\section{Bound state formation in 0-brane collisions\label{D0collisions}}

\begin{figure}
\begin{center}
\includegraphics{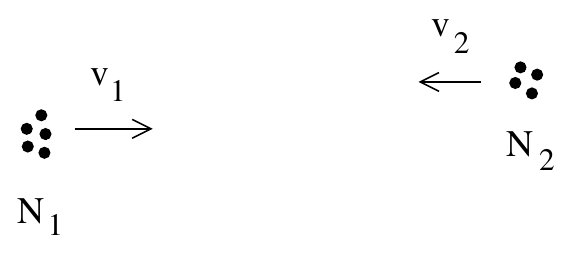}
\end{center}
\caption{Colliding stacks of 0-branes with relative velocity $v$ and impact parameter $b$.\label{collision}}
\end{figure}

Consider colliding two clusters of 0-branes as shown in
Fig.~\ref{collision}.  We'd like to understand whether a bound state
is formed during the collision.  Two mechanisms for bound state
formation have been discussed in the literature.
\begin{enumerate}
\item 
In a perturbative description of D-brane dynamics, open strings can
be produced and lead to formation of a bound state.  This occurs for
impact parameters $b \lesssim \sqrt{v \alpha'}$ \cite{Bachas:1995kx}.
This can be understood as the condition for
violating the adiabatic approximation.  For a review of the
calculation see appendix \ref{appendix:StringCreation}.
\item
At large $N$ and strong coupling the D-brane system has a dual gravitational
description \cite{Itzhaki:1998dd}.  In this description, according
to the hoop conjecture of Thorne \cite{Thorne:1972ji,Gibbons:2009xm,Cvetic:2011vt},
a black hole should form if the two D-brane clusters are contained within their
own Schwarzschild radius.
\end{enumerate}
Our goal is to understand in what regimes these two mechanisms for
bound state formation are operative, and whether they are connected in
any way.

It will be convenient to work in terms of a radial coordinate $U$ with
units of energy, $U = r / \alpha'$.  Here $r$ is the distance between the clusters, $r = \sqrt{b^2 + v^2 t^2}$.  The 't Hooft coupling of the
M(atrix) quantum mechanics is $\lambda = g^2_{\rm YM} N$, which in
string and M-theory units can be expressed as
\be
\lambda = g_s N / \ell_s^3 = R^3 N / \ell_{11}^6\,.
\ee
Here $g_s$ is the string coupling, $\ell_s$ is the string length, $R$
is the radius of the M-theory circle, and $\ell_{11}$ is the M-theory
Planck length.  The mass of a single D0-brane is
\be
m_0 = {1 \over g_s \ell_s} = {1 \over R}\,.
\ee

\subsection{Perturbative string production\label{BoundstateD0}}

We work in the center of mass frame, with momenta
\be
p_1 = {N_1 \over R} \, v_1 \qquad p_2 = {N_2 \over R} \, v_2 \qquad p_1 + p_2 = 0
\ee
We consider a fixed total energy $E$, which determines the
asymptotic relative velocity $v$.
\begin{eqnarray}
\nonumber
&& \quad \quad \, {1 \over 2} \, {N_1 \over R} \, v_1^2 + {1 \over 2} \, {N_2 \over R} \, v_2^2 = E \,, \\
\label{AsympVelocity}
&& \Rightarrow \quad v = v_1 - v_2 \sim \left({N E R \over N_1 N_2}\right)^{1/2} 
= \left({ \lambda E  l^{4}_s \over N_1 N_2}\right)^{1/2}  \,.
\end{eqnarray}
In terms of the $U$ coordinate, the asymptotic relative velocity is
\be
\dot{U} = \left({\lambda E \over N_1 N_2}\right)^{1/2} \,.
\ee
As reviewed in appendix \ref{appendix:StringCreation}, open string production sets in when
\be
U \sim \sqrt{\dot{U}} = \left({\lambda E \over N_1 N_2}\right)^{1/4} \,.
\ee
Note that the radius at which open strings are produced depends on how
we split the total D-brane charge.  The radius is minimized when $N_1
= N_2 = N/2$, which gives the minimum radius for open string
production as
\be
\label{D0openstringbound}
U_0 \sim \left({\lambda E \over N^2}\right)^{1/4} \,.
\ee
This is the case which is interesting for matching to supergravity.

There are some checks we should perform to make sure this perturbative
result is valid.  As discussed in \cite{Becker:1997xw}, the effective
action has a double expansion in $\lambda / U^3$ and $\dot{U}^2 /
U^4$.  The expansion in powers of $\lambda / U^3$ is the Yang-Mills
loop expansion, which is valid provided $U_0 > \lambda^{1/3}$. From
(\ref{D0openstringbound}) this requires
\begin{equation}
\label{PerturbativeCondition}
E > N^2 \lambda^{1/3}
\end{equation}
At the critical point where the loop expansion breaks down, $U_0 \sim \lambda^{1/3}$,
the inequality (\ref{PerturbativeCondition}) is saturated. 

The expansion in powers of $\dot{U}^2 /U^4$ is the derivative
expansion, which is valid when $\dot{U}^2 < U^4$.  Note that the
derivative expansion breaks down at the point where open strings are
produced.  Up to this point, {\it i.e.\ }for $U > \sqrt{\dot{U}}$, one
can trust the two-derivative terms in the effective action, which
means the asymptotic velocity (\ref{AsympVelocity}) is a good approximation to the
actual velocity.\footnote{As we will see, this is not the case
in the supergravity regime.}  So the only condition for the validity
of the perturbative description of open string production is
(\ref{PerturbativeCondition}).

\subsection{Bound state formation in gravity\label{BoundgravityD0}}

At large $N$ the M(atrix) quantum mechanics has a dual gravitational description at
strong coupling, meaning for $U < \lambda^{1/3}$.  So let's imagine
the 0-brane clusters approach to within this distance, and study
whether a bound state can form.

At first, one might think a bound state could form via open string
production.  As noted in \cite{Itzhaki:1998dd}, the metric factors
cancel out of the Nambu-Goto action, and even in the supergravity
regime the mass of an open string connecting the two clusters of
D-branes is $m_W \sim U$.  The adiabatic approximation breaks down,
and these open strings should be produced, if $\dot{U} / U^2 > 1$.
However this velocity cannot be attained in the regime where
supergravity is valid, since it violates the causality bound
\cite{Kabat:1999yq,Silverstein:2003hf}.  This can be seen in the probe
approximation, where the DBI action for a probe is (see, for example, \cite{Iizuka:2001cw})
\be
S = {1 \over g^2_{\rm YM}} \int dt \, 
{U^7 \over \lambda} \left(1 - \sqrt{1 - {\lambda \dot{U}^2 \over U^7}}\right)
\ee
Thus causality bounds the velocity of the probe,
\be
{\lambda \dot{U}^2 \over U^7} < 1\,.
\ee
Rather remarkably, the probe has to slow down significantly as $U
\rightarrow 0$.  In any case, in the supergravity regime we have
${\dot{U}^2 \over U^4} < {U^3 \over \lambda}$, and since ${U^3 \over
\lambda} < 1$ at strong coupling, open strings can never be produced.

This means black hole formation is the only way to form a bound state
in the supergravity regime.  Since open string production is ruled
out, we reach the sensible conclusion that the formation of a horizon
is a purely gravitational closed-string process.  The hoop conjecture
states that a black hole will form if the energy $E$ is contained
within its own Schwarzschild radius.  For a 10-dimensional black hole
with $N$ units of 0-brane charge, the Schwarzschild radius is
\be
\label{D0Schwarzschild}
U_0 = \left({\lambda^2 E \over N^2}\right)^{1/7}
\ee
This 10-D supergravity description is only valid if the curvature and
string coupling are small at the horizon, which requires
\be
\lambda^{1/3} N^{-4/21} < U_0 < \lambda^{1/3}
\ee
For smaller $U_0$ one must lift to M-theory; for larger $U_0$ the
M(atrix) quantum mechanics is weakly coupled. At the outer radius where 
the supergravity approximation breaks down, $U_0 \sim \lambda^{1/3}$, 
eq.~(\ref{D0Schwarzschild}) tells us that $E \sim N^2 \lambda^{1/3}$. 

\subsection{Correspondence point\label{D0correspondence}}

We've found that open string production is only possible at weak
coupling, while black hole formation can only occur within the bubble
where supergravity is valid.  One could ask if the two phenomena are
smoothly connected.  Is there a correspondence point where both
descriptions are valid?

From the perturbative point of view, the transition happens when the
condition (\ref{PerturbativeCondition}) is saturated, $E = N^2
\lambda^{1/3}$.  In this case open strings are produced, but at a
radius $U_0 \sim \lambda^{1/3}$ where the system is just becoming
strongly coupled.

From the supergravity point of view, the transition happens when the
energy of the black hole is $E = N^2 \lambda^{1/3}$, corresponding
to a Schwarzschild radius $U_0 \sim \lambda^{1/3}$.  In this case the
black hole fills the entire region where supergravity is valid.

This suggests that open string production and black hole formation are
indeed continuously connected.  Since the transition between the two
descriptions happens when the curvature at the horizon is of order
string scale,
\be
\alpha' R \sim  (\lambda/U^3)^{-1/2} \sim 1\,,
\ee
this is an example of the correspondence principle of Horowitz and
Polchinski \cite{Horowitz:1996nw}.  Note that for a given
black hole energy, one can view the condition of being at the
correspondence point, $E =  N^2
\lambda^{1/3}$, as fixing the total 0-brane charge,
\be
N = \left({E^3 \ell_s^3 \over g_s}\right)^{1/7} \,.
\ee

\section{D$p$-brane collisions\label{Dpcollisions}}

In this section we generalize our 0-brane results and consider
D$p$-branes wrapped on a $p$-torus of volume $V_p$.  We first record
some general formulas then analyze particular cases.

The Yang-Mills coupling is $g^2_{\rm YM} = g_s / \ell_s^{3 - p}$ and
the 't Hooft coupling is $\lambda = g^2_{\rm YM} N$.  In terms of $U =
r/\alpha'$, the effective dimensionless 't Hooft coupling is
\be
\lambda_{\rm eff} = {\lambda \over U^{3-p}}\,.
\ee
The Yang-Mills theory is weakly coupled when $\lambda_{\rm eff} < 1$.
It has a dual gravitational description at large $N$ when $\lambda_{\rm eff} > 1$
\cite{Itzhaki:1998dd}.

Imagine colliding two stacks of wrapped D$p$-branes at weak coupling,
with a fixed energy density $\epsilon$ as measured in the Yang-Mills
theory.  The mass of a wrapped $p$-brane is $V_p / g_s \ell_s^{p+1}$,
so in the center of mass frame the relative velocity is
\be
\dot{U} = \left({\lambda \epsilon \over N_1 N_2}\right)^{1/2} \,.
\ee
Open string production sets in when
\be
\label{p-braneU}
U \sim \sqrt{\dot{U}} \sim \left({\lambda \epsilon \over N_1 N_2}\right)^{1/4} \,.
\ee
The radius at which open strings are produced depends on how we divide
the total D-brane charge.  The radius is minimized by setting $N_1 =
N_2 = N/2$, which gives the minimum radius for open string production as
\be
U_0 \sim \left({\lambda \epsilon \over N^2}\right)^{1/4} \,.
\ee
This is the case which is interesting for comparison to supergravity.

Just as for 0-branes, open string production is not possible in the
supergravity regime.  The DBI action for a probe brane is
\be
S = {1 \over g^2_{\rm YM}} \int d^{p+1}x \, {U^{7-p} \over \lambda}
\left(1 - \sqrt{1 - {\lambda \dot{U}^2 \over U^{7-p}}}\right)
\ee
Thus the causality bound is $\dot{U}^2/U^4 < U^{3-p} / \lambda =
1/\lambda_{\rm eff}$ \cite{Kabat:1999yq}, which rules out open string
production (at least in the probe approximation).  Instead we have the
process of black hole formation, with a horizon radius $U_0 = (g_{\rm
YM}^4 \epsilon)^{1/(7-p)}$ \cite{Itzhaki:1998dd}.

\noindent
Further analysis depends on the dimension of the branes.

\goodbreak
\noindent
\underline{$p = 0,1,2$}

For $p < 3$ the Yang-Mills theory is weakly coupled when $U >
\lambda^{1/(3-p)}$ and has a dual gravitational description when $U <
\lambda^{1/(3-p)}$.  Thus open string production is possible at large
distances, while black hole formation is possible at small distances.
The correspondence point, where the two descriptions match on to each
other, occurs when
\begin{eqnarray*}
&& \epsilon = N^2 \lambda^{1 + p \over 3 - p} \\
&& U_0 = \lambda^{1 / (3 - p)}
\end{eqnarray*}
At this energy density open string production occurs just as the
Yang-Mills theory is becoming strongly coupled.  From the supergravity
perspective, the resulting black brane fills the entire region in
which supergravity is valid.

\goodbreak
\noindent
\underline{$p = 3$}

In this case the Yang-Mills theory is conformal and dual to ${\rm
AdS}_5 \times S^5$ \cite{Maldacena:1997re}.  The 't Hooft coupling
is dimensionless.  For $\lambda \lesssim 1$ open string production is
possible, while for $\lambda \gtrsim 1$ black holes can form.  The
two descriptions match on to each other at the correspondence point
$\lambda = 1$.  Note that, unlike other values of $p$, the
correspondence point is independent of the energy density
$\epsilon$.

As a test of this idea, note that the radius at which open strings
form is
\be
U_0 = (\lambda \epsilon / N^2)^{1/4}
\ee
while for $p = 3$ the horizon radius is
\be
U_0 = (g^4_{\rm YM} \epsilon)^{1/4}
\ee
These two expressions for $U_0$ agree when $\lambda = 1$.  This suggests that
the process of open string production for $\lambda \lesssim 1$
smoothly matches on to black hole formation for $\lambda \gtrsim 1$.

\goodbreak
\noindent
\underline{$p = 4,\,5,\,6$}

For $p > 3$ the Yang-Mills theory is strongly coupled in the UV and
has a dual supergravity description (modulo some subtleties
described in \cite{Itzhaki:1998dd}).  In the IR the Yang-Mills theory
is weakly coupled.  Black hole production is possible in the
supergravity regime, where $U > \lambda^{1/(3-p)}$, while open string
production is possible for $U < \lambda^{1/(3-p)}$.  The
correspondence point where the two descriptions match is at
\begin{eqnarray}
\label{epsilonNlambdaDprelation}
&& \epsilon = N^2 \lambda^{1 + p \over 3 - p} \\
&& U_0 = \lambda^{1 / (3 - p)}
\end{eqnarray}

\section{Parametric resonance in perturbative SYM\label{sect:parametric}}

In this section we study the evolution of a bound state formed at weak
coupling by open string creation.  We show that the number of open
strings increases exponentially with time due to a parametric
resonance in the gauge theory.  For simplicity we consider $0$-brane
collisions; the generalization to D$p$-branes is straightforward and
will be mentioned in \S \ref{sect:Dptimescale}.

Suppose a cluster of $N_1$ incoming 0-branes collides with a stack of $N_2$ coincident 0-branes at rest.
We assume weak coupling but do not require large $N$.
In the collision suppose $n$ open strings are produced.  These open strings produce
a linear confining potential, so the system will begin to oscillate.  The conserved total energy is
\be
E = {1 \over 2} m v^2 + n \tau x
\ee
Here we're adopting a non-relativistic description, appropriate to the form of the D0-brane quantum mechanics, while
$m$ is the mass of the incoming 0-branes, $v$ is their
velocity, $n$ is the number of open strings created, $\tau = 1/2 \pi \alpha'$ is
the fundamental string tension, and $x$ is the length of the open
strings.  The period of oscillation is
\be
\Delta t = 4 \left({m \over 2}\right)^{1/2} \int_0^{E/n\tau} {dx \over \sqrt{E - n \tau x}} \sim {\sqrt{mE} \over n \tau}
\ee
So up to numerical factors, the frequency of oscillation is
\be
\Omega = {n \tau \over \sqrt{mE}}
\ee
while the amplitude of oscillation (the maximum value of $x$) is
\begin{equation}
\label{YMsize}
L = {E \over n\tau}
\end{equation}

We introduce this as a classical M(atrix) background by setting $X^i = X^i_{\rm cl} + x^i$ where
\be
X^1_{\rm cl} = \left(\begin{array}{cc}
L \sin \Omega t  \, \mathbb{1}_{N_1} & 0 \\
0 & 0
\end{array}\right) \qquad
X^2_{\rm cl} = \cdots = X^9_{\rm cl} = 0 \,,
\ee
We have decomposed the $N \times N$ matrix into blocks; $ \mathbb{1}_{N_1} $ is the $N_1 \times N_1$ unit matrix. 
Expanding to quadratic order in the fluctuations, the M(atrix) Lagrangian\footnote{We are setting $2\pi\alpha' = 1$ and $A_0 = 0$.}
\be
{\cal L}_{\rm YM} = {1 \over 2 g^2_{\rm YM}} {\rm Tr} \left(\dot{X}^i \dot{X}^i + {1 \over 2} [X^i,X^j][X^i,X^j]\right)
\ee
reduces to
\be
{\cal L}_{\rm YM} = {1 \over 2 g^2_{\rm YM}} {\rm Tr} \left(\dot{x}^1 \dot{x}^1\right)
+ {1 \over 2 g^2_{\rm YM}} \sum_{i = 2}^9 {\rm Tr} \left(\dot{x}^i \dot{x}^i + [x^i,X^1_{\rm cl}][x^i,X^1_{\rm cl}]\right)
\ee
Note that the potential for $x^1$ vanishes.  We also have the Gauss constraint associated with setting $A_0 = 0$, namely
\be
\sum_i [X^i,\dot{X}^i] = 0
\ee
To quadratic order this reduces to $[X^1_{\rm cl},\dot{x}^1] = [\dot{X}^1_{\rm cl},x^1]$
which only constrains $x^1$.  The simplest solution is to set $x^1 = 0$.

\begin{figure}
\begin{center}
\includegraphics[width=7.5cm]{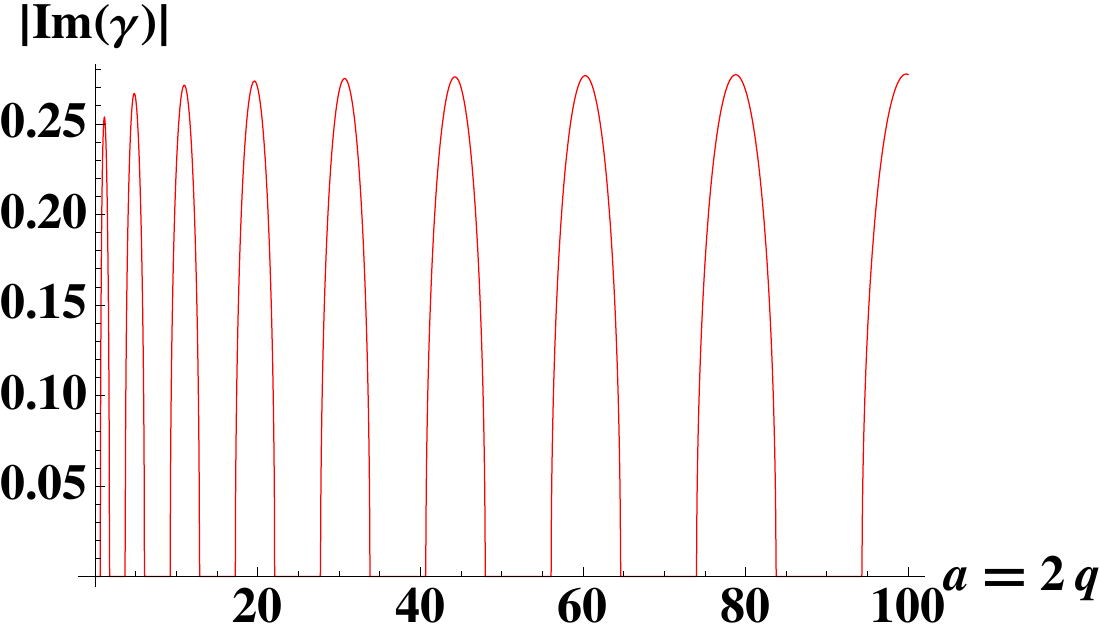}
\end{center}
\caption{The imaginary part of the Mathieu characteristic exponent as a function of $a = 2q$.\label{gamma}}
\end{figure}

To study the remaining degrees of freedom we decompose
\be
x^i = \left(\begin{array}{cc}
a^i & b^i{}^\dagger \\
b^i & c^i
\end{array}\right)
\ee
where $a^i$ is an $N_1 \times N_1$ matrix, $b^i$ is an $N_1 \times N_2 $ rectangular matrix and $c^i$ is an $N_2 \times N_2$ matrix.  We will often
suppress the index $i = 2,\ldots,9$.  To quadratic order the $a$ and $c$ entries have trivial dynamics, 
since $[x^i,X^1_{\rm cl}]$ does not involve $a$ and $c$. 
On the other hand, the equation of motion for $b$ is
\be
\label{Mathieu}
\ddot{b} + L^2 \sin^2(\Omega t) \, b = 0
\ee
Defining $s = \Omega t$ this reduces to Mathieu's equation,
\be
{d^2 b \over ds^2} + (a - 2q \cos 2s) \, b = 0
\ee
with the particular values $a = 2q = L^2 / 2 \Omega^2$.
Mathieu's equation admits Floquet solutions
\be
b(t) = e^{i \gamma \Omega t} P(\Omega t)
\ee
where $P(\cdot)$ is a periodic function with period $\pi$.  As a
function of $a$ and $q$ there are intervals where $\gamma$ has a
negative imaginary part and the solution grows exponentially.  These
intervals correspond to band gaps in the Bloch interpretation of
Mathieu's equation.  The imaginary part of $\gamma$ is plotted as a
function of $a = 2q$ in Fig.~\ref{gamma}.  There are clearly many
intervals where the solution is unstable, with a typical exponent
$\vert {\rm Im} \gamma \vert \sim 0.25$.

This instability corresponds to an exponential growth in
the number of open strings present.  Note that in our case\footnote{Restoring units, we would have $L^2 \rightarrow
L^2 \tau^2$ in (\ref{Mathieu}) and $a = 2q \sim m E^3 / n^4 \tau^2$ in (\ref{a2q}).}
\be
\label{a2q}
a = 2q \sim m E^3 / n^4
\ee
After the initial collision the energy $E$ in the oscillating background will decrease as the system begins to thermalize,
while the number $n$ of open strings gets larger.  So we expect the value of $a$ to decrease with time.  This means the
system will scan across the different instability bands available to it.

To summarize, we have found that the oscillating background resulting from a 0-brane collision is unstable.  The $16 N_1 N_2$
real degrees of freedom contained in $b^i$ for $i = 2,\ldots,9$ behave as parametrically-driven oscillators.  Their amplitude grows
exponentially, on a timescale
\begin{equation}
\label{tYM}
t_{\rm YM} \sim 1/\Omega \sim \sqrt{mE}/n\tau
\end{equation}
Here $m$ is the mass of the $N_1$ incoming 0-branes, $E$ is the total energy of the system, $n$ is the number of open strings present in the
off-diagonal block $b$ and $\tau$ is the fundamental string tension.

\section{Comparison of timescales\label{comparison}}

We compare the timescale associated with parametric resonance to the
quasinormal modes of a black hole.  We consider parametric resonance
for D0-branes in \S \ref{sect:D0timescale}, generalize to D$p$-branes
in \S \ref{sect:Dptimescale}, and compare to quasinormal modes in \S
\ref{quasinormalsection}.

\subsection{0-brane parametric resonance\label{sect:D0timescale}}

As we saw in \S \ref{sect:parametric}, the timescale for parametric
resonance is determined by the period of oscillation.  In a 0-brane
collision this is given by
\begin{equation}
\label{tYMsecond}
t_{\rm YM} \sim 1/\Omega \sim \sqrt{mE}/n\tau
\end{equation}
For $N_1$ incoming D0-branes the mass is $m = N_1/R$, where $R = g_s
l_s$ is the radius of the M-theory circle.  Also $E$ is the total
energy of the system, $n$ is the number of open strings and $\tau \sim
1/l_s^2$ is string tension.  We consider the case $ N_1 \sim N_2 \sim
N$, with $N$ large to compare to supergravity. Then the off-diagonal block $b$ contains ${\cal O}(N^2)$ elements,
so as shown in appendix \ref{appendix:StringCreation} ${\cal O}(N^2)$
open strings are created by parametric resonance.

Using $R = g_s \ell_s$, $\tau \sim 1/\ell_s^2$, $n \sim N^2$ and $g_s \sim g^2_{YM} \ell_s^3$ we obtain  
\be
\label{DeltatD0}
t_{\rm YM} \sim \sqrt{\frac{N E}{R}} \frac{1}{n \tau} \sim \frac{\sqrt{E}}{ \lambda^{1/2} N} \,.
\ee
At the correspondence point
\be
E \sim N^2 \lambda^{1/3}
\ee
which means
\be
\label{deltatlambdaD0}
t_{\rm YM} \sim  \lambda^{-1/3} \,. 
\ee
At the correspondence point the timescale for parametric resonance is independent of $N$ and is set by the 't Hooft scale.
As we will see in \S \ref{quasinormalsection}, the same holds true for the quasinormal frequencies of a black hole at the correspondence point.

\subsection{$p$-brane parametric resonance\label{sect:Dptimescale}}

It's straightforward to extend this result to D$p$-branes.
First, the mass of a single D0-brane in the previous section
is replaced by the mass of D$p$-brane wrapped on a volume $V_p$.  So
we should replace
\be
1/R \to {V_p}/{g_s l_s^{p+1}}\,.
\ee
The energy of the incoming D$p$-branes is related to the energy density $\epsilon$ by
\be
E = \epsilon V_p \,.
\ee
The tension of the strings is the same, $\tau \sim 1/{\ell_s^2}$.
So for D$p$-branes, in place of (\ref{tYMsecond}), the oscillation timescale is
\be
t_{\rm YM} \sim \frac{\sqrt{m E}}{n \tau} \, 
\to \,  V_p  \sqrt{\frac{N \epsilon }{g_s l_s^{p+1}}} \, \frac{1}{n \tau} \,.
\ee
The number of open strings $n$ is modified.  As shown in appendix \ref{appendix:StringCreation}, for $N_1 \sim N_2$ and $p \not= 3$,
the number density of open strings at the correspondence point is set by the 't Hooft scale.  Thus
\be
n \sim N^2 {V_p} {\lambda^{\frac{p}{3-p}}} \,.
\ee
Using this together with $g_s  N = g^{2}_{YM} N \ell_s^{3-p} = \lambda \ell_s^{3-p}$ we obtain 
\be
t_{\rm YM} \sim  V_p  \sqrt{\frac{N \epsilon }{g_s \ell_s^{p+1}}} \, \frac{1}{n \tau} \sim \frac{ \lambda^{-\frac{p}{3-p}} \sqrt{\epsilon}}{\lambda^{1/2} N} \,.
\ee
From (\ref{epsilonNlambdaDprelation}) the energy density at the correspondence point is
\be
\epsilon \sim N^2 \lambda^{\frac{1+p}{3-p}}
\ee
so the timescale is
\be
\label{deltatlambdaDp}
t_{\rm YM} \sim \lambda^{-\frac{1}{3-p}}\,.
\ee
Just as for 0-branes, the timescale for parametric resonance is independent of $N$ and set by the 't Hooft scale.

3-branes are a special case since the 't Hooft coupling is dimensionless.  The correspondence point is defined by $\lambda \sim 1$.
As shown in appendix \ref{appendix:StringCreation}, for $N_1 \sim N_2$ the number of open strings at the correspondence point is
\be
n \sim N^2 V_3 U_0^3
\ee
where $U_0$ is the horizon radius of the black brane.  The energy density at the correspondence point is $\epsilon \sim N^2 U_0^4$,
so the parametric resonance timescale is
\be
t_{\rm YM} \sim  V_p  \sqrt{\frac{N \epsilon }{g_s \ell_s^{p+1}}} \, \frac{1}{n \tau} \sim {1 \over U_0}
\ee
Thus for D3-branes the parametric resonance timescale is $1 / U_0$, which also happens to be the inverse temperature of the black brane.

\subsection{Comparison to quasinormal modes\label{quasinormalsection}}

Quasinormal modes for non-extremal D$p$-branes were studied in
\cite{Iizuka:2003ad,Maeda:2005cr} following earlier work on
AdS-Schwarzschild black holes \cite{Horowitz:1999jd}.  The basic idea
is to solve the scalar wave equation in the near-horizon geometry of
$N$ coincident non-extremal D$p$-branes, with a Dirichlet boundary
condition at infinity and purely ingoing waves at the future horizon.
This gives rise to a discrete set of complex quasinormal frequencies,
whose imaginary parts govern the decay of scalar perturbations of the
black hole.  It was found that the quasinormal frequencies are
proportional to the temperature, with a coefficient of proportionality
that was found numerically in \cite{Iizuka:2003ad}.

Recall that the temperature, energy density and entropy density of
these black branes are related to their horizon radius $U_0$ by
\cite{Itzhaki:1998dd,Iizuka:2003ad}
\beas
&& T \sim {1 \over \sqrt{\lambda}} U_0^{(5-p)/2} \\
&& \epsilon \sim {N^2 \over \lambda^2} U_0^{7-p} \\
&& s \sim {N^2 \over \lambda^{3/2}} U_0^{(9-p)/2}
\eeas
Assuming $p \not= 3$, at the correspondence point we have $U_0 \sim \lambda^{1/(3-p)}$ so that
\beas
&& T \sim \lambda^{1 \over 3-p} \\[2pt]
&& \epsilon \sim N^2 \lambda^{p+1 \over 3 - p} \\[2pt]
&& s \sim N^2 \lambda^{p \over 3 - p}
\eeas
These quantities all obey the expected large-$N$ counting, and since the 't Hooft coupling $\lambda$ has units of $({\rm energy})^{3-p}$, these
results could have been guessed on dimensional grounds.  In the special case $p = 3$ the 't Hooft coupling is dimensionless and the correspondence
point is defined by $\lambda = 1$.  At the correspondence point the horizon radius $U_0$ remains arbitrary, with
\beas
&& T = U_0 \\
&& \epsilon = N^2 U_0^4 \\
&& s = N^2 U_0^3
\eeas
Again these results could have been guessed on dimensional grounds.

As we saw in \S \ref{sect:D0timescale} and \ref{sect:Dptimescale} the timescale for parametric resonance is
\be
t_{\rm YM} \sim \left\lbrace
\begin{array}{ll}
\lambda^{-1/(3-p)} \qquad & \hbox{\rm for $p \not= 3$} \\
1/U_0 & \hbox{\rm for $p = 3$}
\end{array}\right.
\ee
For all $p$ this matches the inverse temperature of the black brane, $t_{\rm YM} \sim 1/T$.  Thus at the correspondence point the
timescale for parametric resonance matches the timescale for the decay of quasinormal excitations of the black brane.

\section{Comparison to equilibrium properties\label{sect:equilibrium}}

It's interesting to compare the properties of the bound state as initially formed to the equilibrium properties
of the black hole.  This will show us that, at the correspondence point, very little additional evolution is required to
reach equilibrium -- perhaps just a few $e$-foldings
of parametric resonance will suffice.

First, in a 0-brane collision, note that the total number of open strings produced is $\sim N_1 N_2$.
With equal charges $N_1 = N_2 = N/2$ the number of open strings is ${\cal O}(N^2)$.  At the correspondence
point these strings have a mass $\sim \lambda^{1/3}$, so the total energy and entropy in open strings is
\beas
&& E \sim N^2 \lambda^{1/3} \\
&& S \sim N^2
\eeas
This matches the equilibrium energy and entropy of the black hole, suggesting that black hole formation at the correspondence point
is a simple one-step procedure, in which the open strings that are formed in the initial collision essentially account for the equilibrium properties
of the black hole.  The analogous result for $p$-branes is that the number of open strings at the correspondence point is, for $p \not= 3$,
\be
n \sim N^2 V_p \lambda^{p \over 3-p}
\ee
where we have used (\ref{appendix:n}) and the fact that $U \sim \lambda^{1 \over 3 - p}$.
Since the open strings have a mass $\sim U$, this corresponds to a total energy and entropy in open strings
\beas
&& E \sim N^2 V_p \lambda^{p + 1 \over 3 - p} \\
&& S \sim N^2 V_p \lambda^{p \over 3 - p}
\eeas
which again matches the equilibrium energy and entropy of the black
brane.  This again suggests that the black hole is essentially
fully formed in the initial collision, with very little additional
evolution required to reach equilibrium.\footnote{When $p = 3$ the
matching is $n \sim N^2 V_3 U_0^3$, $E \sim N^2 V_3 U_0^4$, $S \sim
N^2 V_3 U_0^3$.}

Another quantity we can compare at the correspondence point is the
size of the bound state.  At weak coupling, after $n$ open strings
have been formed, the amplitude of oscillation of the resulting bound
state is, from (\ref{YMsize}),
\be
L = \frac{E}{n \tau}
\ee
At the correspondence point for general $p$ we have
\be
E \sim N^2 V_p U_0^{p+1}
\ee
while the initial number of open strings created is
\be
n \sim N^2 V_p U_0^p
\ee
Thus the initial amplitude of oscillation as measured in the $U$ coordinate is
\be
L / \ell_s^2 = E / n \sim U_0
\ee
In other words, the initial oscillation amplitude matches the equilibrium horizon radius of the black brane.
Again this suggests that after the initial collision, only a small amount of additional evolution is required to reach equilibrium.

\section{Shell Collapse\label{gaugeshellcollapse}}

\begin{figure}
\begin{center}
\includegraphics[width=8cm]{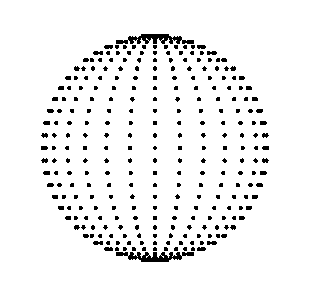}
\end{center}
\caption{A collapsing shell of 0-branes.  Initially the 0-branes are spread uniformly over an $S^8$ with velocities toward the center.\label{collapse}}
\end{figure}

So far we have studied bound state formation in a collision between
two clusters of D-branes, in the geometry shown in
Fig.~\ref{collision}.  Here we study a different initial
configuration, in which $N$ D0-branes are uniformly distributed over a
collapsing spherical shell as in Fig.~\ref{collapse}.  We will see
that the correspondence principle applies and a similar outcome is
obtained in this case.

We consider an initial configuration in which the 0-branes are
uniformly spread over an $S^8$ of radius $U$ in 9 spatial dimensions.
The 0-branes are localized but uniformly distributed over the sphere,
with velocities directed toward the center.  Intuitively we argue as
follows.  Since the total volume of the sphere scales as $U^8$, each
0-brane occupies a volume $\sim U^8/N$, and the distance between
nearest-neighbor 0-branes scales as $U / N^{1/8}$.  This means virtual
open strings connecting nearest-neighbor 0-branes are quite light,
with a mass $\sim U / N^{1/8}$ that goes to zero at large $N$.
However the typical open string is much heavier, with a mass $\sim U$
that is independent of $N$.  We expect these typical open strings to
dominate the bound-state formation process, and therefore expect to
have a well-defined correspondence point at large $N$.

\begin{figure}
\begin{center}
\includegraphics[width=7cm]{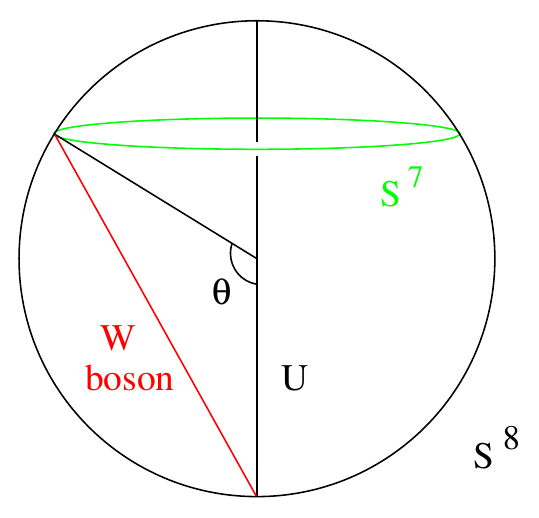}
\end{center}
\caption{The 0-branes are spread over an $S^8$ of radius $U$.  The green $S^7$ has radius $U \sin \theta$ and the red
W boson has length $2 U \sin \theta/2$.\label{sphere8}}
\end{figure}

\begin{figure}
\begin{center}
\includegraphics[scale=0.65]{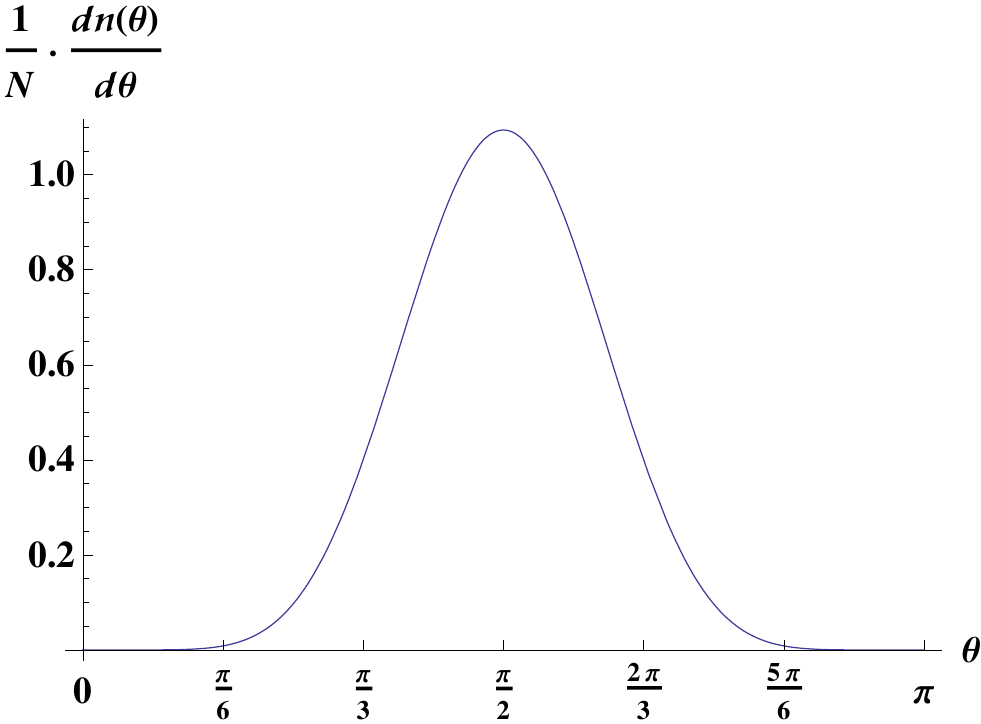}
\includegraphics[scale=0.65]{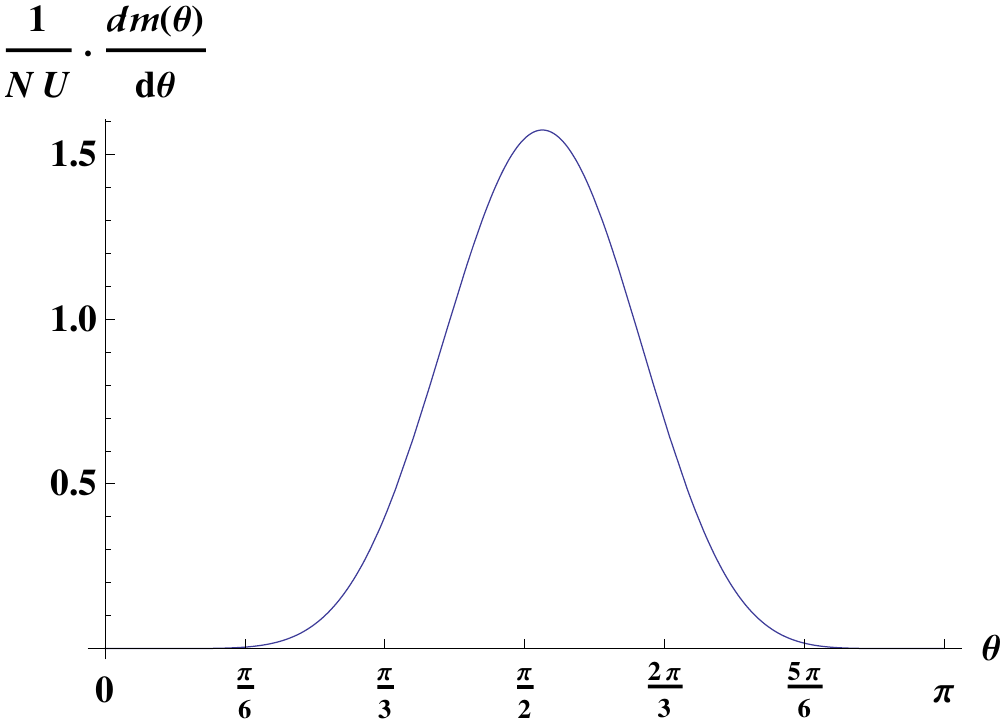}
\end{center}
\caption{On the left, the W-boson number density ${1 \over N} {dn \over d\theta}$.  On the right,
the W-boson mass density ${1 \over NU} {dm \over d\theta}$.\label{distribution}}
\end{figure}

To argue this in more detail, it is useful to consider a 0-brane
located at the south pole and study the number of virtual open strings as a
function of the angle $\theta$ to the other 0-brane.  See
Fig.\ \ref{sphere8}.  The number of distinct open strings $dn$ in the
interval $(\theta,\theta + d \theta)$ is
\be
dn = \frac{N}{\frac{32 \pi^4 U^8}{105}} \times {\pi^4 \over 3} ( U \sin \theta)^7 \times U d \theta
\ee
The first factor $N / \big(\frac{32 \pi^4 U^8}{105}\big)$ is the
number density of 0-branes on the $S^8$, the second factor ${\pi^4
\over 3} (U \sin \theta)^7$ is the volume of an $S^7$ located at an
angle $\theta$ from the south pole.  Thus the number density of open
strings is
\be
{dn \over d\theta} = \frac{35}{32} N \sin^7 \theta
\ee
We can also find the mass density of open strings ${dm \over d\theta}$.  Since an open string subtending an angle
$\theta$ has a mass $2 U \sin \theta / 2$, this is given by
\be
{dm \over d\theta} = {dn \over d\theta} \cdot 2 U \sin {\theta \over 2} = \frac{35}{16} N U \sin^7 \theta \sin \frac{\theta}{2}
\ee
The W-boson number density ${1 \over N} {dn \over d\theta}$ and mass density ${1 \over NU} {dm \over d\theta}$ are plotted in Fig.~\ref{distribution}. 

As can be seen in the figure, there are light open strings at large
$N$.  However the number of these strings is tiny, since ${dn \over d\theta}
\sim \theta^7$ at small angles.\footnote{This is due to the fact that
the 0-branes are spread on an $S^8$.  The distribution would be less
sharply peaked in lower dimensions, with ${dn \over d\theta} \sim
\theta^{d-1}$ on an $S^d$.}  Most of the W-bosons are concentrated
around $\theta = \pi/2$. Therefore a spherical shell is basically the
same as having W-bosons distributed in the interval $\theta_0 < \theta
< \pi - \theta_0$, where $\theta_0$ is determined by the fraction of
0-branes pairs we neglect.  For example, if we neglect ${dn \over d\theta} \le
10^{-7} N$, then $\theta_0 \sim 0.1$.  Since the masses of the W-bosons
near $\theta = \pi/2$ are all $O(U)$, we can simply approximate the
entire W-boson spectrum by taking $m_W \sim U$.

We now consider what happens when we give the shell of 0-branes some velocity toward the origin.
The analysis is almost identical to the colliding clusters considered in \S \ref{D0collisions}.
Given $N$ D0-branes with total energy $E$, the asymptotic relative velocity is  
\bea
&& E \sim {\rm mass} \times v^2 \sim \frac{N}{R} v^2 \nonumber \\
& \Rightarrow & v \sim \left( \frac{E R}{N} \right)^{1/2}
= \left( \frac{E \lambda l^4_s}{N^2} \right)^{1/2}
\eea
In terms of the $U$ coordinate, this becomes 
\be
\dot{U} = \left( \frac{E \lambda}{N^2} \right)^{1/2} 
\ee
This matches the result in \S \ref{D0collisions} for $N_1 = N_2 \sim N$. 
Since the W-boson masses are concentrated around $m_W \sim U$, open string production again sets in when
\be
U \sim \sqrt{\dot U} \sim  \left( \frac{E \lambda}{N^2} \right)^{1/4} 
\ee
At the correspondence point, where the effective gauge coupling becomes order one, we have
\be
U \sim \lambda^{1/3}
\ee
and therefore
\be
E \sim N^2 \lambda^{1/3} \,.
\ee
Just as in \S \ref{D0collisions}, this matches the radius and energy energy of a black hole at the correspondence point.

\section{Conclusions\label{sect:conclusions}}

In this paper we studied D-brane collisions.  We argued that the
process of open string creation, which leads to formation of a D-brane
bound state at weak coupling, smoothly matches on to a process at
strong coupling, namely black hole formation in the dual supergravity.
The transition happens at an intermediate value of the coupling, given
by the correspondence principle of Horowitz and Polchinski.  The size
of the bound state, the timescale for approaching equilibrium, and the
thermodynamic properties of the bound state all agree between the two
descriptions.  The latter agreement happens quickly, which suggests
that the bound state is formed by the initial collision in a
near-equilibrium configuration.

We considered two types of initial configurations, namely colliding
clusters of wrapped D$p$-branes and a collapsing shell of D0-branes.
The main difference between the two configurations was that the shell
had a tail of light open strings which we argued could be neglected.
In fact, this distinction between the two configurations is somewhat
artificial, since with somewhat more generic initial conditions the
0-branes which make up the clusters could have some small random
relative velocities.  One would then expect a bit of open string
production within the clusters, which would put the two examples on
much the same footing.

In the examples we studied the powers of $N$ were fixed by large-$N$ counting,
so at the correspondence point there was
essentially only a single length scale in the problem, namely the 't
Hooft scale (for $p \not= 3$) or the horizon radius (when $p = 3$).
In a sense this guaranteed the matching between perturbative gauge
theory and gravity results, just on dimensional grounds.  To explore
this further it would be interesting to study multi-charged black
holes, or to deform the background in a way which introduces another
length scale, and ask whether there is still a simple transition
between perturbative worldvolume dynamics and black hole formation.

A step in this direction would be to consider 0-brane collisions but
with $N_1 \not= N_2$.  In this case, as we saw in \S
\ref{sect:equilibrium}, the matching between perturbative gauge and
gravity results must be more complicated, because the energy and
entropy in open strings that are created in the initial collision do
not match the equilibrium energy and entropy of the black hole.  This
means further dynamical evolution is required before the bound state
reaches equilibrium.  It would be interesting to study this, perhaps
by going beyond the linearized approximation made when studying
parametric resonance in \S \ref{sect:parametric}.  There are several
related interesting examples to consider, for example a situation in
which several concentric layers of shells are collapsing.

Another direction would be to use the present results to better
understand the microstructure of black holes.  The picture that
emerges, that a black hole is a thermal bound state of D-branes and
open strings, is reminiscent of the fuzzball proposal
\cite{Mathur:2005zp}.  However the real question, relevant for understanding
firewalls \cite{Almheiri:2012rt} or the energetic curtains of \cite{Braunstein:2009my}, is whether this thermal state could
be a dual description of the interior geometry of the black hole.

\bigskip
\goodbreak
\centerline{\bf Acknowledgements}
\noindent
DK is grateful to David Berenstein for numerous discussions on this
topic.  S.R.\ thanks Justin R.\ David for discussions.  This work was
supported in part by U.S.\ National Science Foundation grants PHY-0855582 and PHY-1125915 and by grants from
PSC-CUNY.  The research of SR is supported in part by Govt.\ of India Department of Science and Technology's research grant under scheme DSTO/1100 (ACAQFT).

\appendix
\section{String production in a D-brane collision\label{appendix:StringCreation}}

We review the process of open string production in a D-brane
collision, following \cite{Bachas:1995kx,Douglas:1996yp}.

Consider colliding two 0-branes with relative velocity $v$ and impact
parameter $b$.  Setting $2 \pi \alpha' = 1$, the virtual open strings
connecting the two 0-branes have an energy or frequency $\omega =
\sqrt{v^2 t^2 + b^2}$.  As long as this frequency is changing
adiabatically open strings will not be produced.  The adiabatic
approximation breaks down when $\dot{\omega}/\omega^2 \gtrsim 1$.  The
peak value of this quantity is $\dot{\omega}/\omega^2 \sim v/b^2$ when
$vt \sim b$, so (restoring units) open strings are produced for $b
\lesssim \sqrt{v \alpha'}$.  In terms of the radial coordinate $U = r / \alpha'$,
where $r$ is the distance between 0-branes,
the energy of an open string is $m_W = U/2\pi$.  So the adiabatic approximation
breaks down and open strings are produced when $\dot{U}/U^2 \sim 1$.\footnote{In
principle we should distinguish between the asymptotic relative velocity $\dot{U} = v/\alpha'$
and the actual time-dependent value $\dot{U} = {v \over \alpha'} \, {vt \over \sqrt{b^2 + v^2t^2}}$.
But at $vt \sim b$ this distinction can be ignored.}

Now consider colliding two $p$-branes wrapped on a torus of volume
$V_p$, with relative velocity $v$ and impact parameter $b$ in the
transverse dimensions.  Consider a virtual open string that connects
the two $p$-branes and has momentum $k$ along the $p$-brane
worldvolumes.  Setting $2 \pi \alpha' = 1$, this virtual open string
has an energy or frequency
\[
\omega = \sqrt{k^2 + v^2 t^2 + b^2}
\]
If $k = 0$ then the condition for open string production is just what
it was for 0-branes, $b \lesssim \sqrt{v}$.  Having non-zero $k$
increases $\omega$ and suppresses open string production.  Effectively
there is a cutoff, that open strings are produced up to a maximum
momentum $k \sim b \sim \sqrt{v}$.  Restoring units, the maximum
momentum is $k \sim \sqrt{v / \alpha'} = \dot{U}^{1/2}$.  This cutoff
corresponds to a number density of open strings on the $p$-brane
worldvolume
\[
{\hbox{\rm \# open strings} \over \hbox{\rm volume}} \sim \dot{U}^{p/2}
\]
Again these open strings are produced when $\dot{U}/U^2 \sim 1$.

If we collide two stacks of D$p$-branes with charges $N_1$ and $N_2$
respectively, it's easy to estimate the total number of open strings
that are produced.  At weak coupling the individual brane collisions
are independent events.  So for 0-branes the total number of open
strings produced is
\[
n \sim N_1 N_2
\]
while for $p$-branes the total number of open strings produced is
\be
n \sim N_1 N_2 V_p \dot{U}^{p/2}
\ee
or equivalently, in terms of the radius at which open string production takes place
\be
\label{appendix:n}
n \sim N_1 N_2 V_p U^p
\ee

There is, however, an important consistency check on this result: we
need to make sure the incoming D-branes have enough kinetic energy to
produce this number of open strings.  Equivalently, we need to make
sure that the back-reaction of open string production on the
velocities of the D-branes is under control.  Given the number of open
strings (\ref{appendix:n}), the energy in open strings is
\[
E_{\rm string} = n U = N_1 N_2 V_p \left({\lambda \epsilon \over N_1 N_2}\right)^{p + 1 \over 4}
\]
where we have used (\ref{p-braneU}).  On the other hand the kinetic
energy of the incoming branes is
\[
E = \epsilon V_p
\]
Thus the ratio
\be
{E_{\rm string} \over E} = \lambda \left({\lambda \epsilon \over N_1 N_2}\right)^{p - 3 \over 4}
\ee
and the consistency condition $E_{\rm string} / E < 1$ is equivalent to
\[
\lambda \, U^{p - 3} < 1
\]
This is nothing but the condition $\lambda_{\rm eff} < 1$.  Thus at
weak coupling energy conservation does not limit the number of open
strings that are produced and the simple estimate (\ref{appendix:n})
can be trusted.


\providecommand{\href}[2]{#2}\begingroup\raggedright\endgroup

\end{document}